# Chemical Vapor Deposition Synthesized Atomically Thin Molybdenum Disulfide with Optoelectronic-Grade Crystalline Quality

Ismail Bilgin,[1,2] Fangze Liu,[1] Anthony Vargas,[1] Andrew Winchester,[3,4] Michael Ka Lun Man,[4] Moneesh Upmanyu,[5] Keshav M. Dani,[4] Gautam Gupta[2], Saikat Talapatra,[3,4] Aditya D. Mohite*[2] and Swastik Kar*[1,6]

[1]Department of Physics,  Northeastern University, Boston, Massachusetts 02115, United States
[2]Materials Synthesis and Integrated Devices, Los Alamos National Laboratory, Los Alamos, New Mexico 87545, United States
[3]Department of Physics, Southern Illinois University Carbondale, Carbondale, Illinois 62901, United States
[4]Femtosecond Spectroscopy Unit, Okinawa Institute of Science and Technology Graduate University, Onna, Okinawa 904-0495, Japan
[5]Department of Mechanical and Industrial Engineering, Northeastern University, Boston, Massachusetts 02115, United States
[6]George J. Kostas Research Institute for Homeland Security, Northeastern University, Burlington, Massachusetts 01803, United States

*Authors for Correspondence: amohite@lanl.gov and s.kar@neu.edu

## ABSTRACT

The ability to synthesize high-quality samples over large areas and at low cost is one of the biggest challenges during the developmental stage of any novel material. While chemical vapor deposition (CVD) methods provide a promising low-cost route for CMOS compatible, large-scale growth of materials, it often falls short of the high-quality demands in nanoelectronics and optoelectronics. We present large-scale CVD synthesis of single- and few- layered MoS₂ using direct vapor-phase sulfurization of MoO₂, which enables us to obtain extremely high-quality single-crystal monolayer MoS₂ samples with field-effect mobility exceeding 30 cm²/Vs in monolayers. These samples can be readily synthesized on a variety of substrates, and demonstrate a high-degree of optoelectronic uniformity in Raman and photoluminescence mapping over entire crystals with areas exceeding hundreds of square micrometers. Owing to their high crystalline quality, Raman spectroscopy on these samples reveal a range of multi-phonon processes through peaks with equal or better clarity compared to past reports on mechanically exfoliated samples. This enables us to investigate the layer thickness- and substrate-dependence of the extremely weak phonon processes at 285 cm⁻¹ and 487 cm⁻¹ in 2D MoS₂. The ultra-high, optoelectronic-grade crystalline quality of these samples could be further established through photocurrent spectroscopy, which clearly reveal excitonic states at room temperature, a feat that has been previously demonstrated only on device fabricated with mechanically exfoliated that were artificially suspended across trenches. Our method reflects a big step in the development of atomically thin, 2D MoS₂ for scalable, high-quality optoelectronics.

### TOC Graphic

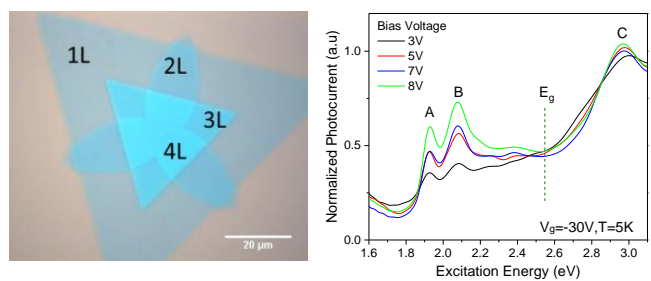

In recent times, there has been a rapidly growing interest in atomically thin, layered, or 2D materials beyond graphene.[1] Confined in a 2D-plane, these materials demonstrate a range of exciting properties such as emergent photoluminescence,[2] anomalous lattice vibrations,[3] control of valley polarization using optical helicity,[4, 5] excitonic dark states,[6] and strong light-matter interactions at heterojunctions,[7] and have immense potential for next-generation transistors with extremely high on-off ratios,[8] photodetectors,[9] catalysis[10] and other applications. Amongst the variety of materials being investigated, atomically thin molybdenum disulfide (2D-MoS$_2$) has emerged as one of the most promising candidates for low cost, flexible and highly sensitive next-generation nanoelectronic [8, 9, 11, 12] and optoelectronic devices such as photodetectors, photovoltaics, and light emitting diodes.[13-17] Unlike graphene, 2D-MoS$_2$ is a true semiconductor with layer-thickness-dependent band gap that becomes direct at the monolayer limit (E$_g$ ~1.85 eV).[2, 18] The direct band gap in monolayer MoS$_2$ is extremely attractive both for light-emission applications, as well as for photocurrent-based applications since it is possible to obtain valley polarization of excitons using circularly polarized light.[4, 5] Moreover, the sheet resistance of atomically thin MoS$_2$ can be modulated significantly by either applying a gate voltage or by incident light, making it valuable for diverse electronic and optoelectronic applications. These properties overcome the fundamental drawbacks of graphene-based electronic devices that are limited by a lack of band-gap, wavelength-independent weak optical absorption of about 2.3%,[19] and extremely fast non-radiative recombination of photogenetated electron-hole pairs,[20] and hence enable a wide range of applications in nanoelectronic and optoelectronic devices such as photoemitters, phototransistor and photodetectors. To successfully utilize these attractive attributes of 2D-MoS$_2$, and to integrate them into existing optoelectronic platforms require scalable synthesis of large-area, high quality single-crystal (possibly monolayer) samples with uniform electronic/optoelectronic responses. At this point, simultaneously accomplishing these conditions for 2D-MoS$_2$ still appears to be an open quest, especially for optoelectronics. Conventionally, the highest quality samples are usually those obtained by mechanical exfoliation of bulk MoS$_2$. Field-effect transistors (FETs) using exfoliated MoS$_2$ exhibit high saturation currents >100 μA/μm with a factor of difference >10$^6$ between the ON state and OFF state, and the field effect mobilities between 0.1 and 10 $cm^2V^{-1}s^{-1}$,[8, 13, 21] Mechanically exfoliated $MoS_2$ is also sensitive to a certain range of the visible spectrum, making it appealing for photodetection.[22] However, certain low-intensity optoelectronic processes are difficult to observe even in exfoliated samples. For example, it has been recently shown that demonstration of excitonic states in photocurrent spectra of 2D-$MoS_2$ using photocurrent spectroscopy in exfoliated samples require utmost care in device fabrication, including suspending the layer to remove deleterious effects of the substrate, *in-situ* annealing, and performance of measurements in low temperatures.[23] Moreover, although first order-Raman spectra of 2D-$MoS_2$ has been studied in extreme detail in this system, low-intensity multi-phonon processes and higher-order Raman peaks have been reportedly observed, in a very limited manner, only in exfoliated samples.[24-26]

While mechanically exfoliated samples have helped elucidate the diverse potentials of 2D-$MoS_2$, their poor yield, uncontrolled and irregular sample homogeniety make them unsuitable for any practical applications. In contrast, chemical vapor deposition (CVD) has potential for large-scale, low-cost manufacturing 2D-$MoS_2$ of uniform layer-thicknesses over macroscopic sizes-scales, and such samples are widely viewed as ideal for integration with current CMOS platform. Recently, CVD growth of monolayer $MoS_2$ has been obtained by sulfurization of Mo[27] and $MoO_3$[28-30] powders as a precursor. The latter recipe has become a standard for the CVD growth of monolayer $MoS_2$. These samples also possess field-effect mobilities between 1-10 $cm^2V^{-1}s^{-1}$. While most reports focus on electronic mobility as a parameter to quantify sample quality, the demonstration of sensitive optoelectronic processes where electron-photon and electron-phonon coupling processes are not destroyed by defects, disorder, or impurities, remains largely unexplored. The observation of rare optoelectronic processes can be thought of as a necessary and perhaps more stringent measure of the high crystalline quality of these samples, especially for optoelectronic applications.

Detailed study of the sulfurization of $MoO_3$ has shown that the conversion of $MoO_3$ to $MoS_2$ involves an intermediate step, during which $MoO_3$ is first partially reduced to $MoO_2$,[31, 32] which then sulfurizes under appropriate conditions to produce $MoS_2$. Wang *et al.* have reported a method to separate these two steps whereby $MoO_3$ powder was first thermally evaporated, reduced using sulfur, and re-condensed to obtain $MoO_2$ micro-crystals, which in the second step were sulfurized layer-by-layer to obtain 2D-$MoS_2$ crystals.[33] Finally, a PMMA-assisted cleavage step was needed to separate $MoS_2$ from the underlying $MoO_2$ crystal and transfer onto other substrates. In this context, the question arises whether there is any need or benefit of using $MoO_3$ as the initial precursor, instead of $MoO_2$, which is a more stable oxide[32]. The relevance of this question increases when one carefully scrutinizes the reaction pathway of $MoS_2$ synthesis. Li *et al.*[32] postulated that the reduction of $MoO_3$ follows the step: $MoO_3$ + (x/2) S → $MoO_{3-x}$ + (x/2) $SO_2$, followed by a sulfurization of $MoO_{3-x}$, *i.e.* $MoO_{3-x}$ +(7-x)/2 S → $MoS_2$ + (3-x)/2 $SO_2$. In the ideal scenario, x=1 results in intermediate compound being $MoO_2$, and each removed oxygen atom participates in forming $SO_2$. In the relatively less-ideal conditions within a CVD chamber, however, incomplete reaction may result in uncontrolled amounts of $MoS_{2-y}O_y$ phase in the atomically thin $MoS_2$ crystal.[34] Further, as the $MoO_3$ is sulfurized to produce $MoS_2$ crystals, these crystals provide additional binding sites for the oxygen to chemisorb on. [35] Oxygen, especially at high temperatures is also a well-known etchant for 2D materials[36, 37] which suggests that at the very least, any nascent oxygen could potentially create defects/vacancies during growth. Beyond structural and chemical effects, the presence of oxygen can also have important impacts on the electronic properties of atomically thin $MoS_2$. For example, adsorbed oxygen has been shown to significantly reduce sheet conductivity as well as mobility of atomically thin $MoS_2$.[38, 39] It appears that avoiding a two-step reaction process that could at least partially decrease the possibility of incomplete reactions, as well as reduce the possible deleterious effects of oxygen may provide atomically thin $MoS_2$ samples with improved structural, chemical and electronic properties.

In this work, we present the synthesis and characterization of higly crystalline single- and few-layered samples of 2D-$MoS_2$ with large-area single domains, by direct sulfurization of $MoO_2$ in the vapor phase. We believe that the direct vapor-phase sulfurization of $MoO_2$, represented by the single-step chemical reaction $MoO_2$ +3S → $MoS_2$ + $SO_2$, circumvents the need for any intermediate chemistry, and leads to samples of higher quality. We show that controllable

mono-, and few-layer $MoS_2$ can be grown on a variety of subtrates using $MoO_2$ as a precursor. The high degree of uniformity of these samples could be verified by direct and differential Raman mapping, as well as through PL mapping. Multi-layer samples that have well-defined geometric shapes enable us to observed layer-thickness-dependent evolution of a range of multi-phonon and higher-order Raman processes that have previously never been observed in CVD-grown $MoS_2$ samples. More importantly, we have been able to observe two new *layer-thickness-dependent* Raman modes that have previously never been reported in *any* $MoS_2$ samples. Back-gated measurements in FET configuration shows high saturation current and an ON/OFF ratio $\approx 10^6$. Finally, a direct reflection of the high crystalline quality of our samples is the observation of the clear peaks (even at room temperature) in energy-resolved photocurrent measurements that correspond to excitonic states in single-layer $MoS_2$. Exciton-generated peaks in photocurrent spectroscopy has never been observed before in CVD-grown 2D-$MoS_2$, and has been recently reported to occur only in artificially-suspended, mechanically exfoliated 2D-$MoS_2$ samples. This spectroscopic tool allow us to clearly resolve the two primary excitonic peaks in 2D-$MoS_2$, unlike photoluminescence peaks that strongly overlap. This provides a powerful method to investigate the independent and comparative evolution of each peak as a function of carrier energy (tuned using voltage and temperature). These investigations provide a fundamental platform for understanding the optoelectronic properties of large-area, CVD-grown 2D-$MoS_2$.

**RESULTS AND DISCUSSION**

Single and few-layered $MoS_2$ samples were fabricated by direct sulfurization of $MoO_2$ (see SI for details) as a precursor instead of commonly used $MoO_3$ and Mo, and without any treatment to the precursor[30, 33] or substrate[28]. This latter feature is quite useful, since pretreatment of substrates has been shown to be detrimental to the intrinsic properties of 2D materials.[40] Figure 1 outlines a range of different types of samples that can be readily obtained by our method, and on a range of substrates. In a separate work, we have now also developed a method for substrate-free direct synthesis of $MoS_2$ across micron-size apertures using the same precursor, which has been reported elsewhere.[41] Figure 1(a) shows optical and SEM images of single-layer

MoS$_2$ grown on a variety of substrates including amorphous, crystalline, transparent, and conductive substrates, with typical sample edge-sizes ranging from 10-50 μm. It is indeed quite remarkable that despite their variety, this single-step synthesis method requires no pre-treatment of either the precursor or the substrates. The ability to readily synthesize these samples on a range of substrates makes it convenient to utilize these as-prepared samples to be directly implemented in a variety of nanoelectronic, optoelectronic, and catalytic (including photo/electro-catalytic) applications. Moreover, these as-grown samples (free of transfer-induced contamination) on such a variety of substrates are also potentially attractive for a range of advanced metrology tools such as STM, TEM, ARPES, and various optical spectroscopies, to name a few.

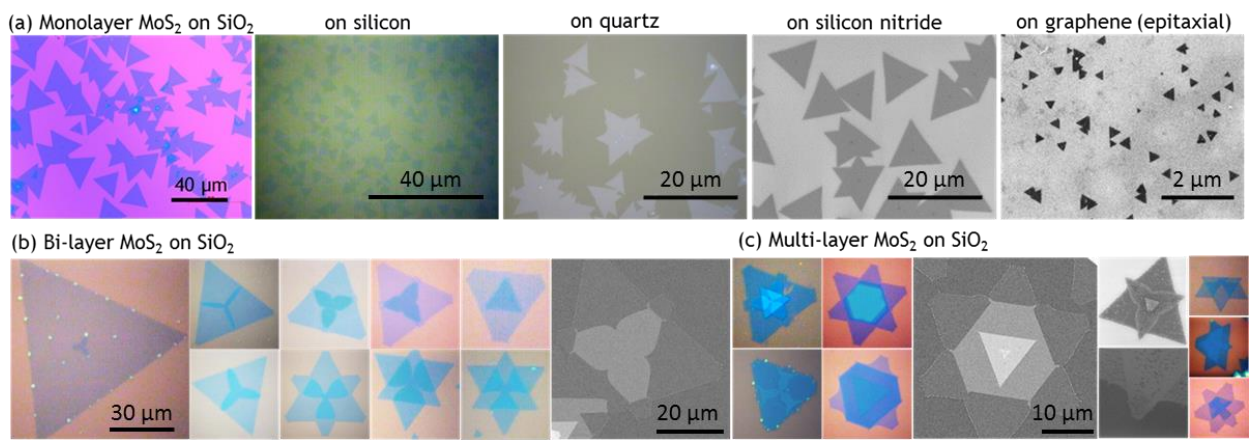

**Figure 1**: (a) Typical example of monolayer MoS$_2$ samples grown using MoO$_2$ precursor, on different substrates. Colored images are optical, and B/W ones are SEM images. (b) Typical optical and SEM images of bi-layer MoS$_2$ samples fabricated on SiO$_2$. Various states of initiation and growth of the second layer have been represented by these images. The final image in (b) is an SEM image showing intricate details of the bilayer structure. (c) Optical and SEM images of a range of multi-layer samples at various stages of growth and with various shapes and morphologies.

Changing the synthesis conditions triggers second layer growth vertically on top of the first layer. Figure 1(b) outlines a typical array of samples at various stages of second-layer growth on SiO$_2$ (see SI for additional details). In several cases, we also see growth of third and higher layers (Figure 1(c)). Growth of multiple layers is unsurprising as MoS$_2$ exhibits a natural tendency to grow as a 3D crystal[42]. Of interest is the growth morphology of the second and higher layers which are varied and in stark contrast to the growth of the first layer. The substrate-independent morphology of the first layer suggests that it is likely a Wulff shape composed of either low energy edges (near-equilibrium shape), or kinetically-limited edges (far-from-equilibrium kinetic shape). The three-fold symmetric shape is consistent with past reports on $MoS_2$ growth using oxygen-rich precursors[29], indicating that the thermodynamic and kinetic conditions at the growing edges in our direct sulfurization process are not dramatically altered.

The concurrent growth of the second layer indicates that its rate of areal increase is fast relative to the first layer. As confirmation, the second layer coverage always increases over time. The fast growth rate suggests that the accumulation of precursor molecules is mediated by the underlying first layer. As disucussed earlier, in contrast to sulfurization of $MoO_3$, which involves the intermediate formation of $MoO_{3-x}$, the ability to grow these samples using a direct vapor-phase sulfurization of $MoO_2$ may imply a faster growth kinetics. This may at least be partially responsible for the rapid formation of the second layers. The shape is again three-fold symmetric but dendritic, and the primary arms usually grow towards center of the edges triangular first layer, *i.e.* the second layer is rotated by $(\pi/3+\theta)$ with respect to the first layer, where $\theta$ is the relative stacking-based lattice rotation between the two layers that lowers the energy of the two layers[43]. The morphology of an early stage seed is shown in Figure 1(b). The presence of the first layer also modifies the anisotropy in the shape. To see this, note that as the areal coverage of the second layer increases, the growth morphology becomes progressively less dendritic (right, Figure 1(b)). The primary arm pinches off and the shape asymptotes towards a rotated triangular (top right row, Figure 1(b)), and in other cases the dendritic tip flattens out in a ``wine glass'' fashion as it approaches the edge of the underlying first layer (bottom right row, Figure 1(b)). The SEM images in Figure 1b and 1c, which appear to provide evidence of grain boundaries, suggest that it may be the presence of such grain boundaries that initiate the formation of the dendritic morphologies, although further studies will be required to arrive at a firm conclusion.

Taken in toto, the dendritic growth is reminiscent of diffusion-limited growth (DLG)[44], likely due to slow diffusion of precursor molecules to the growing edges. The accelerated growth can lead to a transition in the growth dynamics, and this has been observed in oxygen-controlled growth of single layer graphene domains[44].  In the case of bilayer growth as is the case here, our results indicate the morphological relation between the two layers, suggesting a strong influence of crystal symmetry of the first layer on the kinetic shape. The anisotropy can be a combined effect of first-layer mediated diffusion of precursor adaptors and/or attachment-detachment kinetics at the growing edge (aggregation limited growth, ALG)[44]. Similar kinetic effects shape the growth of the third and higher layers.

Interestingly, as the edges of the two layers begin to interact and compete for the attachment of precursor molecules, we see growth of another dendrite from the first layer. This is quite common in the first two layers, and results in an overall star-shaped morphology of the first layer.  The dendrites appears to have a different lattice orientation and are therefore separated by the grain boundaries; localized contamination at these  boundaries is visible in the SEM images in Figures 1(b) and 1(c). These new domains can result from overgrowth of the second layer mediated by the edge of the first layer, or a net effect of inter-edge interaction that starves the growth of the second layer by facilitating nucleation and edge-by-edge growth of new domains at the edges of the first layer. We conclude that the system provides a interesting platform for controlling the shape and morphology of atomically thin crystals.

Since 2D materials are known to show strong layer-thickness dependent properties, the ability to controllably synthesize these samples with various layer-thicknesses, that too on a variety of substrates (not shown) opens up several significant layer-thickness-dependent nanoelectronic and optoelectronic investigative directions as well as applications. Next, we establish the quality of these samples using  Raman and Photoluminescence Spectroscopy and mapping, and present results of these Raman spectroscopy as a function of layer thickness and on different substrates.

Figure 2a shows the AFM topographical image at the edge of a multi-layer sample. Cross-sectional step-height measurement from various regions of a partially unfinished top-layer, as shown on the right, establishes that the single-layer thickness of $MoS_2$ ranges from 0.5-0.8 nm, the variation could be attributed to the roughness of the underlying $SiO_2$ substrate. Figure 2b shows the dominant Raman peaks measured from a sample with very well-defined and characterized 1,2,3, and 4-layer regions, shown in the inset, measured with an incident laser of wavelength = 488 nm (2.54 eV). The two most dominant peaks seen correspond to the first-order $E^1_{2g}$ and $A_{1g}$ modes at at the Γ-point of the hexagonal Brillouin zone of 2D-$MoS_2$, for different layer thickness values. We note that the low intensity of the $A_{1g}$ peak is due to the polarized configuration of the Raman system.[45] The $A_{1g}$ peak can be fully recovered by adding a half waveplate (see Figure S6 in Supporting Information). As reported earlier[28, 30, 46], the $E^1_{2g}$ peak position ($\omega[E^1_{2g}]$) red-shifts, while the $A_{1g}$ peak position ($\omega[A_{1g}]$) blue-shifts with increasing layer thickness, resulting from the growing influence of inter-layer coupling on the electron-phonon processes in 2D-$MoS_2$. Figure 2c shows the variation of $\Delta = \omega[A_{1g}]-\omega[E^1_{2g}]$ , the difference in peak-position between the $E^1_{2g}$ and $A_{1g}$ modes, as function of layer thickness, measured from multilayered samples grown on $SiO_2/Si$ and quartz substrates. In samples grown on $SiO_2$ substrates, we see ~20% variation in the value of $\Delta$ in monolayers, ~10% in bilayer samples, and almost negligible variations in 3 and 4 layered samples. In contrast, there is about 3-5% variation in $\Delta$ for all samples grown on quartz, independent of their layer-thickness. This leads us to believe that substrate-induced effects (*e.g.* doping due to impurities and trapped charges, substrate-sample interactions, strain and even sample background noise as seen in figure 2h) may play a measurable role in the measured optoelectronic properties of 2D-MoS$_2$, and these effects vary substantially from substrate to substrate. Indeed, these interactions could also be responsible for the differences in the behavior of other Raman spectroscopic features, as discussed later.

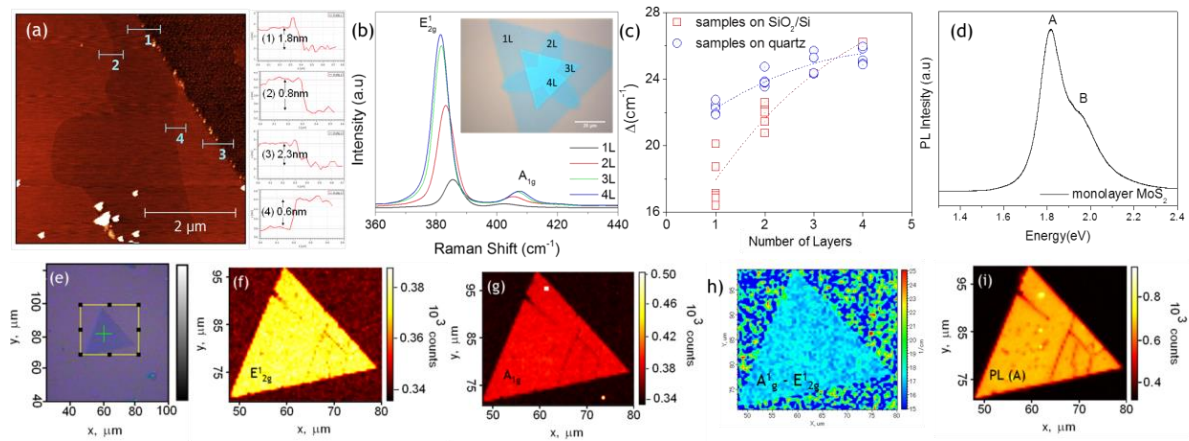

**Figure 2:** (a) AFM image of a few-layered sample, with incomplete top-layer, such that the top-layer thickness can be measured using the cross-section step height, as shown on the side. (b) Raman spectra from regions of various layer-thickness from a multilayer sample (Inset: optical image), showing the evolution of the dominant $E^1_{2g}$ and $A_{1g}$ peaks for different layer-thicknesses. (c) Variation of the peak position separation, $\Delta = \omega[A_{1g}]-\omega[E^1_{2g}]$, as a function of layer-thickness in samples grown on $SiO_2/Si$ and quartz substrates. The spread in datapoint for each layer corresponds to 4-6 datapoints that were collected from different samples. (d) Typical Photoluminescence (PL) spectrum obtained from a monolayer $MoS_2$ sample, showing the A and B excitonic peaks. (e) Optical image of a monolayer sample, which was used to perform Raman and PL mapping, as shown in (f) Raman map of $\omega[E^1_{2g}]$, (g) Raman map of $\omega[A_{1g}]$ (h) Raman map of $\omega[A_{1g}]-\omega[E^1_{2g}]$ and (i) PL map of the Excitonic A-peak.

Monolayer MoS$_2$ is a direct band-gap semiconductor,[2, 18] and when e-h pairs in monolayer MoS$_2$ are excited with photons of energy larger than this gap size of E$_g$~1.85 eV, they recombine radiatively, giving rise to a photoluminescence (PL) specrtrum that contains two distinct peaks. These corresponds to transitions from the lowest energy point of the spin-orbit-split lowest conduction bands to the highest point of the valence bands, commonly designated as A and B peaks in monolayer MoS$_2$. Figure 2d shows the PL spectrum of a typical monolayer MoS$_2$ sample, where the A and B peaks are found to be centered around 1.82 eV and 1.94 eV, respectively. The Raman and PL peaks obtained in these samples have been used to demonstrate the optoelectronic homogeneity of these samples. Figure 2e shows a typical monolayer MoS$_2$ sample on which extensive Raman and PL mapping were performed. Figures 2f-2h show pseudo-colored maps, where the colors represent the values of the direct ($\omega[E^1_{2g}]$ and $\omega[A_{1g}]$) and differential ($\Delta=\omega[A_{1g}]-\omega[E^1_{2g}]$) peak positions for the $E^1_{2g}$ and $A_{1g}$ peaks. The excellent uniformity of the direct peak position maps (other than the few streaks corresponding to wrinkles in these 2D materials) reflects the high degree of uniformity within each monolayer crystal. The differential peak position map shows a uniformly distributed background noise which can at least be partially responsible for the variation of $\Delta$ seen in figure 2c. Although the uniformity could not be clearly established in the differential map owing to these large background fluctuations, similar degree of uniformity could also be found from the PL map of the A peak, as seen in figure 2i. With the help of these measurements, we establish the uniform nature of the quality of these samples. We next present more evidence of the extremely high quality of these CVD-grown samples with the help of previously unreported, multi-phonon processes in the Raman spectra as a function of layer-thickness.

Figure 3a shows the low-intensity region of the same curves plotted in figure 2b, revealing a rich range of first order and multi-phonon processes. Such large number of Raman peaks have previously not been reported in CVD-grown samples, and were confirmed to be present in all our samples tested, including samples grown on different substrates. A number of peaks could be labelled with the help of past reports on bulk or exfoliated $MoS_2$ samples, and are labeled in black.[24-26, 47] A broad peak is observed around ~454 $cm^{-1}$, which is attributed to second order longitudinal acoustic phonons near the M-point of the Brilluoin zone (2LA(M)) of $MoS_2$, although the first-order LA(M) phonon[26] was absent when measured using $\lambda$=488 nm ($E_{ph}$ = 2.54 eV) in our samples. The broadness of this peak is consistent with past reports that it is a combination of three[25] to five[47] closely placed peaks. The LA(M) phonons also combine with $A_{1g}$ phonons to give sum ($A_{1g}$+LA(M) at ~635 cm$^{-1}$) and difference ($A_{1g}$-LA(M) at ~184 cm$^{-1}$) peaks. A peak at ~595 cm$^{-1}$ could be either due to a combination of $B_{1u}$(M)+$A^2_{2u}$(M) phonons, or a second-order $2E_{2u}(\Gamma)$ peak; most likely a combination of the two. A range of higher-order peaks were also clearly observed between the range of 700-850 cm$^{-1}$.[47]

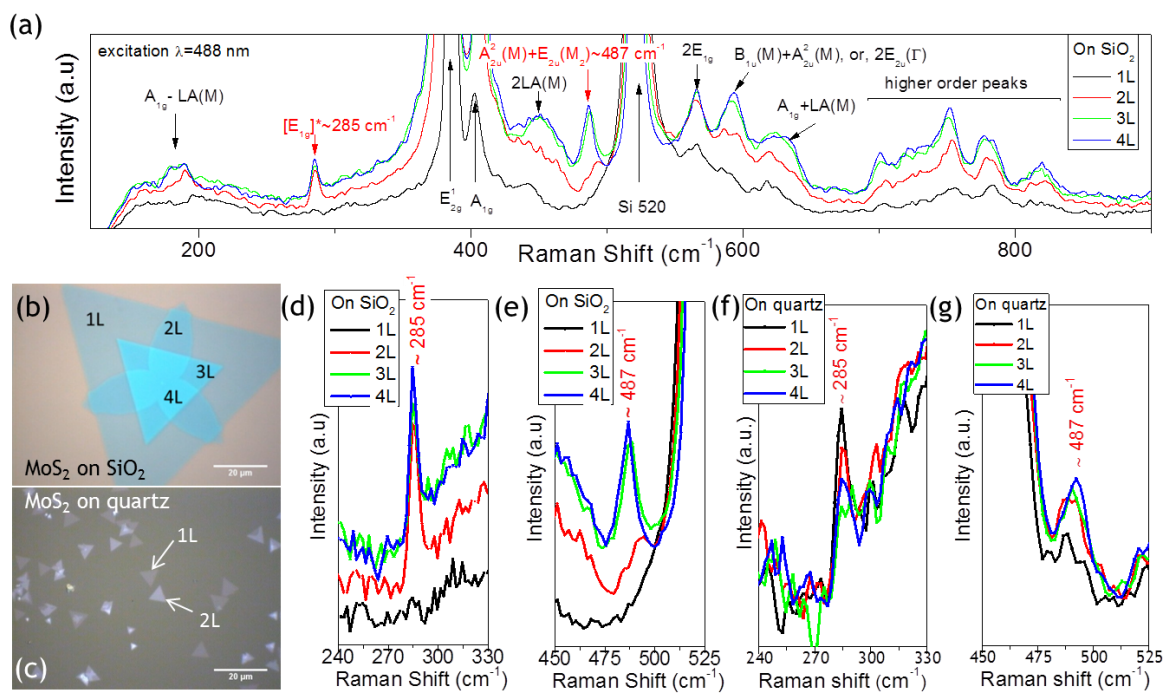

**Figure 3**: (a) Typical survey scan of several first-order and multi-phonon processes identified in monolayer and few-layered 2D-$MoS_2$ samples synthesized on $SiO_2$/Si. The intensities of most of these processes are much weaker than the two dominant processes $E^1_{2g}$ and $A_{1g}$, and have been labeled as shown. The two labels in red are previously unreported peaks (see text). (b) The few-layered CVD-grown sample synthesized on $SiO_2$/Si (same as figure 2b Inset), and (c) representative few-layered CVD-grown samples synthesized on quartz that were used to investigate the layer-thickness-dependence of the two new peaks, details of which are shown in (d)-(g). (d), (e) Absence of Raman peak at ~285 $cm^{-1}$ and ~487 $cm^{-1}$ in monolayer samples that appear in samples with more layers, as seen in 2D-$MoS_2$ grown on 300 nm of $SiO_2$ on silicon. (f)-(g) Averaged layer-thickness variation of the same peaks when measured on 2D-$MoS_2$ grown on quartz (see text).

The most interesting features observed in our samples are two previously unreported peaks at ~285 cm$^{-1}$ and ~487 cm$^{-1}$ (during the review process, we became aware of another work that independently observed the ~285 cm$^{-1}$ peak)[48]. Figure 3b and c shows optical images of multi-layered samples on $SiO_2$ (300 nm)-on-silicon and quartz substrates, repectively, and 3d–3g show Raman spectra from layers of different thicknesses near these two new peaks from these and similar samples. In samples grown on $SiO_2$/Si substrates the peak at ~285 cm$^{-1}$ was found to be absent in all monolayer samples, but appeared strongly in samples with more layers (as seen in figure 3d). The origin of this peak can be traced to a the $E_{1g}$ phonons near the $\Gamma$-point,[47] and has been previously reported in few-layered WSe$_2$ samples.[49] According to Scheuschner *et al.*[48] single-layer, few-layer and bulk MoS$_2$ belong to different point groups, with the $E_{1g}$, $E_g$ and E' modes in bulk, even layers and odd layers, respectively, all originating from the E'' mode in single-layer MoS$_2$. The E'' mode requires a scattering geometry with a z-component to be observable which is difficult to realize in backscattering in the single layer,[26] which is consistent with its absence in 1L sample on $SiO_2$/Si. When measured on samples grown on quartz, these peaks were difficult to discern from the background level, and required significantly higher data-averaging. Fig. 3 f show the variation of this peak as a function of layer-thickness, where each spectrum shown is an average of 5 spectra measured on different samples. We find, surprisingly, that the peak at ~285 cm$^{-1}$ shows up for *all* layer-thicknesses, including in monolayer samples. Moreover, within our experimental limits, it appears that the averaged peak-height was strongest in monolayer samples. These observations are in stark contrast to the spectral behavior in samples grown on $SiO_2$. Our data appears to imply that the mono- and few-layer samples interact with quartz in a manner which is different from that with $SiO_2$, and which enables the realization of the E'' mode in the backscattering geometry. Detailed future studies will be required to understand the underlying physics that causes the differences between the layer-thickness-dependence of Raman spectra obtained on samples on these two and possibly other substrates. Interestingly, the position of $E_g$/E′/E″ phonon modes in 2D-MoS$_2$ were found to be resilient against layer-thickening in samples on both substrates, within our experimental resolution of 1 cm$^{-1}$.

The second newly-observed peak at ~487 cm$^{-1}$ appears to be a combination of transverse acoustic phonons $A^2_{2u}$(M) near the M-point and the $E_{2u}$(M$_2$) phonon near the M point with $A_u$ symmetry. The layer-thickness dependent appearance of this peak has been shown in figure 3e (sample on SiO$_2$ substrate) and 3g (sample on quartz substrate). In samples grown on SiO$_2$/Si, this mode was also found to be absent in monolayer samples. While a very weak peak appears in only some bi-layer samples, it appears strongly in samples of higher number of layers. In contrast, evidence of this mode could be found in even monolayer samples of MoS$_2$ grown on quartz, with growing peak intensity as the layers thickened. In this case, we were able observe a a blue shift in the peak position from ~488 cm$^{-1}$ to ~493 cm$^{-1}$ as the layer thickness grew from 1-4. The presence of this new Raman peak in our samples is surprising since it has been predicted[47] to arise only at low temperatures. We note that at this point, the reasons leading to the differences in the layer-thickness variation of this peak in samples grown on different substrates is not clear to us. Nevertheless, at the very least, we believe that our ability to observe these very subtle differences is testament to the high optoelectronic quality of our 2D-MoS$_2$ crystals.

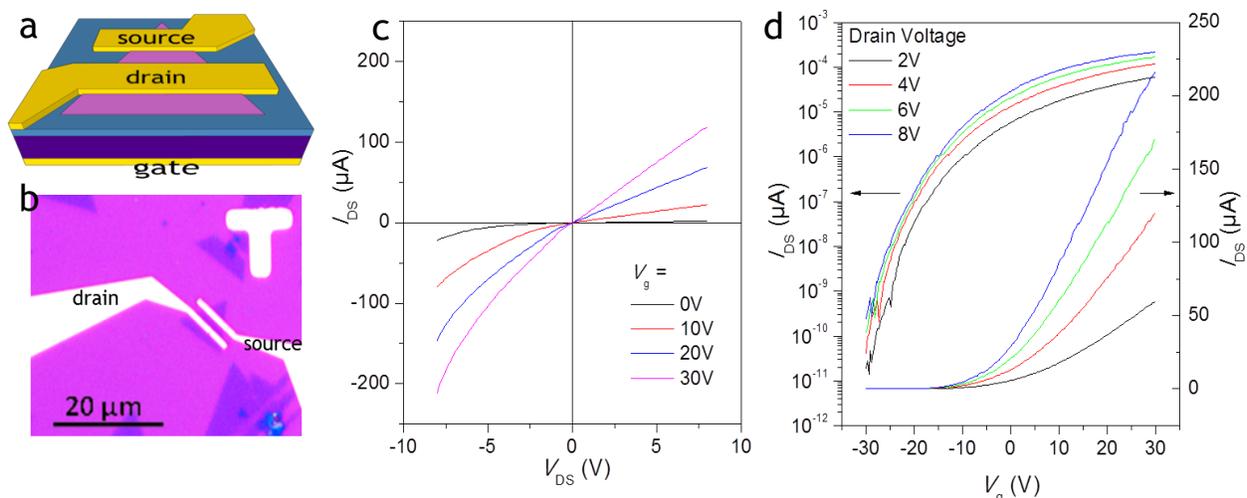

**Figure 4:** (a) Schematic illustration of the single layer $MoS_2$ field effect transistor platform for electrical and photocurrent measurements. (b) Optical image of the FET device. (c) $I_d$-$V_d$ curves measured for different gate voltages in the dark. (d) $I_d$-$V_g$ curve measured for a bias voltage ranging from 2 V to 8 V.

To measure the gate-modulated electrical and optoelectronic properties of single-layer $MoS_2$, as-grown samples were transferred onto d=100 nm thick $SiO_2$/Si substrates (pre-marked to assist lithography) by using a polymer-assisted transfer method. The schematic of a field-effect transistor device structure is shown figure 4a. Electrical contacts were fabricated with electron-beam lithograghy followed by the deposition of 5 nm Ti and 50 nm Au using an e-beam evaporator. Figure 4b shows the optical image of the actual device on which measurements were performed. In Figure 4c, $I_d$-$V_d$ curves are shown for different gate voltages which appear to exhibit low schottky-barrier characteristics. Figure 4d shows the transfer characteristics ($I_d$-$V_g$) of the same device measured at different source-drain voltages. Single layer $MoS_2$ FET devices exhibit a typical n type characteristic with a threshold voltage -10 V, in good agreement with previous reports.[8, 17, 50] The field-effect mobility of this device was $\approx$ 35 $cm^2$/Vs at room temperature for our devices calculated using the equation $\mu$=[d$I_d$/d$V_g$]×[L/W$C_i$$V_d$],[8] where the channel length L=2.4 $\mu$m, channel with W=1.4 $\mu$m, back gate capacitance per unit area $C_i = \varepsilon_r \varepsilon_0/d$ ($\varepsilon_r$=3.9, d=100 nm) and $V_d$=10 V. The mobility of our devices are about 2 times higher than those reported for monolayer CVD grown $MoS_2$ using other precursors[27-30], and comparable to the values of mobilities that are obtained by post annealing the samples[51, 52].The $I_{on}$/$I_{off}$ ratio obtained for $V_d$=8 V is ~$10^6$ for gate voltages in the range of -30 to + 30 V. At the same voltage, the subthreshold swing (S=d$V_g$/d(log$I_d$)) ~ 5 V/dec in good agreement with previous results.[53, 54] Our preliminary electronic characterizations of these devices demonstrate significant improvement over past reports both for CVD-grown as well as mechanically exfoliated monolayer MoS$_2$.

Photocurrent spectroscopy was performed on these devices using a broadband white light source coupled to a monochromator allowing us to tune the wavelength of the incident light from 400 nm – 1200 nm continuously. The output from the spectrometer was coupled to a Vis-NIR multimode fiber and focused onto the device, which was mounted in a Janis cryo probe station. The net photocurrent was calculated by taking the difference between the dark current the total output current with illumination. A measurable photocurrent was only observed when a finite bias was applied between the source and the drain electrodes and its value was found to be strongly bias dependent. This observation helps us eliminate the possibility of any photovoltaic contribution from the electrode-MoS$_2$ junctions. Moreover, the photocurrent is significantly larger (see SI Fig 3) than the dark current at gate voltage less than -10V, where the channel is undoped and results in a sharp decrease of the latter. Below $V_g$ = -30 V, the channel was assumed to be "intrinsic", and the photocurrent spectra were measured under these intrinsic conditions by scanning the wavelength of incident light and recording DC photocurrent as a function of applied source drain bias and temperature.

Figure 5a shows the room-temperature photocurrent measured as a function of excitation wavelength for a single layer MoS$_2$ FET device. The measured photocurrent from the MoS$_2$ devices was normalized with the incident power at the end of the fiber optic used for photo excitation. The incident power for each wavelength was calibrated using the spectral response (Amp/Watt) of a standard Si photodiode placed at the same distance from the fiber optic as the device. A series of peaks are clearly seen at wavelengths λ=646 nm, =608 nm, and =440 nm (corresponding to photon energy values of 1.92 eV, 2.04 eV and 2.82 eV, respectively) that are labelled as A, B and C recpectively. The position of the peaks are in a good agremement with previous experimental reports on absorption spectroscopy[2, 18, 55], photocurrent spectra on measured on mechanically exfoliated and suspended samples,[23] and light scattering experiments on $MoS_2$ samples on gold substrate[56]. We attribute the peaks at 1.92 eV and 2.04 eV to the A-and B- excitons as a result of valence band splitting at the K point due to spin-orbit (SO) splitting and the absence of inversion symmery.[2, 18, 57] Owing to the clear splitting of the two excitonic peaks, it was possible to accurately determine the energy difference of ~120 meV between the position of A and B peaks, which corresponds to the energy due to SO coupling and compares well with absorption measurements reported previously.[2, 18] Since A and B are excitonic states, it is expected that the observed excitonic peaks would be electric field dependent. The presence of an external field would aid the dissociation and separation of bound electron-hole pairs thereby increasing the photocurrent efficiency. Indeed, as illustrated in figure 5b, there is an overall increase in the magnitude of the photocurrent peaks corresponding to the A and B excitons. The peak labeled C is associated with van-Hove singularities in the DoS where arises between K and Γ points at the brillouin zone[18, 23, 58] consistent with measurements performed by Klots *et al*. By taking first derivativative of the photocurrent spectra with respect to energy, we could extract the continuum band gap energy value at room temperature $E_g$ ≈ 2.4 eV, as illustrated in Fig. 5b inset. This value is consisted with  the experimentally calculated single layer $MoS_2$  devices.[23, 59]  We further obtain an exciton binding energy of 480 meV by substracting the band-edge from the position of the excitonic peak A at room temperature ($E_b$=$E_g$-$E_A$). Analysis of the photocurrent amplitudes of excitonic states at room temperature indicates an increase in the amplitude ratio of A/B (see Fig. 5c), which appear to imply that at least at room temperature, the A exciton can dissociate more efficienctly than the B exciton and contributes more to the net photocurrent. To our knowledge, such clear and detailed information regarding the excitonic properties using photocurrent spectroscopy has previously not been reported on CVD-grown MoS$_2$, which we believe is an indication of the high optoelectronic-grade quality of our samples.

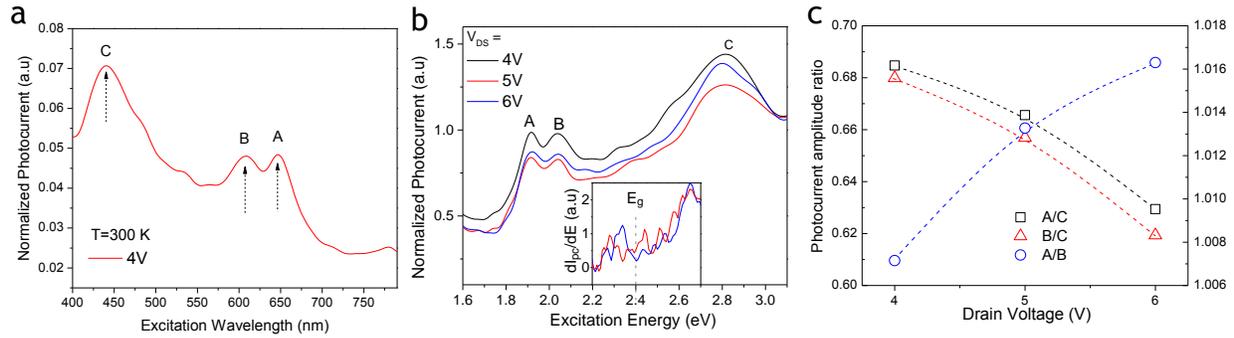

**Figure 5**. Excitonic states in monolayer $MoS_2$ investigated by photocurrent spectroscopy at room temperature. (a) Excitonic states A, B, and C characterized by the corresponding peaks in the photocurrent spectrum ($V_{ds}$= 4 V). (b) Bias dependence of photocurrent spectra showing evolution of excitonic peaks ($V_g$ = -42 V). Inset: First derivative of the photocurrent spectra with respect to energy, from which the contiuum band gap energy of single layer $MoS_2$ is ≈ 2.4 eV. (c) Bias dependence of the photocurrent amplitude ratios of the A-,B-, and C- excitonic peaks at room temperature.

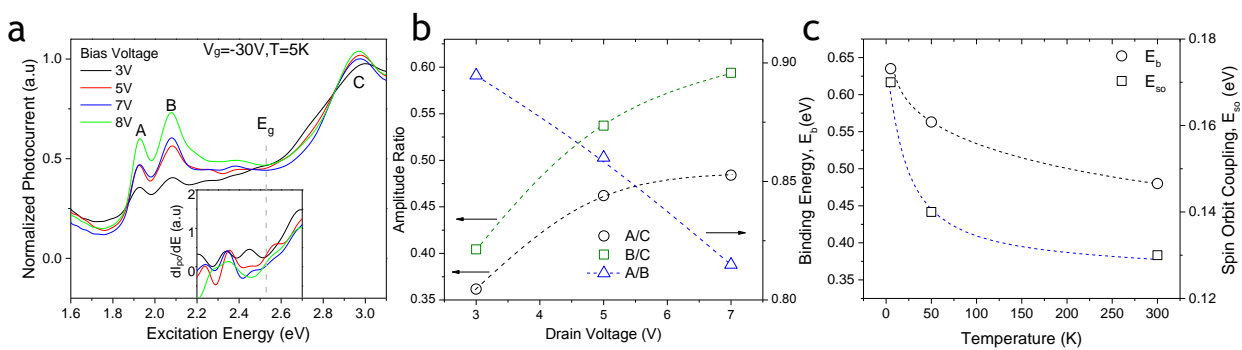

**Figure 6.** Photocurrent spectra of single layer MoS$_2$ at low temperature (5 K): (a) PC spectra measured for different bias voltages at V$_g$=-30 V, 5 K. The inset: Band gap calculation of single layer MoS2 at low temperature. (b) Relative photocurrent amplitudes of A-, B-, & C- peaks. (c) Binding and spin orbit coupling energies for single layer MoS$_2$ at different temperatures.

Further information regarding the optoelectronic energetics could be obtained by performing photocurrent spectroscopy at low temperatures. Fig 6. shows the photocurrent spectra of the same device measured at 5 K with increasing V$_{sd}$. It can be clearly inferred that the photocurrent peaks are sharp and much more well-defined as compared to room temperature. The photocurrent of B exciton becomes slightly larger with respect to A state and we also find that the ratio of A/C and B/C increase with increasing drain-source voltage as illustrated in fig. 6b. The position of the peaks remain unchanged with bias voltages at 5 K. Fig. 6c shows the variation in the exciton binding energy and the SO coupling energy spacing with temperature. We observe a decrease in both of these values as we approach RT. These observations are consistent with observations on devices fabricated with exfoliated MoS$_2$ flakes suspended across a trench.[23] To the best of our knowledge this is the first observation of the excitonic states in the photocurrent spectrum of CVD grown MoS$_2$ devices and attribute this to the high optoelectronic quality of our single-crystalline MoS$_2$ flakes.

**CONCLUSIONS**

In conclusion, a range of higly crystalline monolayer and few-layered $MoS_2$ structures can be controllably grown *via* CVD method by using $MoO_2$ as a source. Using this method, $MoS_2$ structures can be obtained on a broad variety of subtrates that makes it suitable both for fundamental investigations as well as applications development, without the need for any pre-treatment on precursors. While mostly monolayer samples can grow at CVD temperatures ranging from 650 – 850 ℃, we did not find any direct correlation between the samples of relatively better quality and synthesis temperature. For the purpose of standardizing, we have selected the middle of this range, *i.e.* 750 ℃, at which temperature, adequate surface coverage of single crystal flakes within a reasonable growth duration was obtained. High-resolution TEM resolved images of our samples, when grown substrate-free[41] show extremely high crystalline quality in our samples in comparison to those reported earlier[28, 31, 33]. In addition to the commonly observed Raman and PL peaks, sensitive information regarding multi-phonon processes could be clearly observed on these samples, including two previously unreported Raman peaks. The field-effect mobility of these samples are notably higher than those measured in samples synthesized using Mo and $MoO_3$ based precursors. For the first time, we have demonstrated excitonic states in photocurrent spectra at room temperature in CVD-grown samples of $MoS_2$, which was previously thought to be impossible to demonstrate even in exfoliated samples without suspension. We believe that our observations establish that using $MoO_2$ as a source leads to high-quality atomically-thin crystals of $MoS_2$ samples which is a big step towards fundamental research and application development of the optoelectronic devices.

## METHODS

Large area monolayer MoS$_2$ was synthesized by CVD method using MoO$_2$ as a source. Monolayers are synthesized low temperature at 750 °C in quartz tube 1 inch diameter by atmospheric pressure CVD. A 300 nm SiO$_2$/Si substrate is cleaned in acetone, isopropyl alcohol and deionized water and is placed face down alumina boat containing 10 mg MoO$_2$ powder (99% Sigma Aldrich) at the center of the furnace. 20 mg Sulfur powder (99.5% Alfa Aesar) is placed to upstream at the edge of the furnace. The tube is flushed 3 times with Ar carrier gas at room temperature before starting growth. The furnace temperature is first increased to 300 °C and sits there one hour with 100 sccm Argon. Then the temperature ramps up slowly to growth temperature of 750 °C at 3 °C /min with 200 sccm Argon and held there 15 minutes before cool down the room temperature.

Bi –Tri layer MoS$_2$ structures are obtained high temperature at 950 °C . The sulfur powder is placed far from the furnace edge(2.5 inch ). The furnace was heated to 650 °C at 30 °C /min with 150 sccm Ar then increased to 950 °C at 5 °C /min with 200 sccm Ar carrier gas and stay there 10 minutes. The sulfur is started to melt around 900 ℃.

### Device Fabrication

MoS$_2$ flakes on 300 nm Si/SiO$_2$ are transferred to p type (R<0.0015 ohm-cm) 100 nm Si/SiO$_2$ substrate by PMMA transfer method. First, PMMA C4 was spin coated at 4000 rpm for 60 s and baked 135 ℃ for 2:30 min. Then the chip was immersed in 1M KOH solution for an hour . Obtained PMMA and MoS$_2$ film transferred to new substrate. This was followed by acetone and IPA cleaning to remove PMMA residues. FET devices were made on 100 nm Si/SiO$_2$ samples by E-beam lithograpy using PMMA C4 or A4. The electrodes (5 nm Ti/50 nm Au) were deposited by e –beam evaporator with rate deposition 1 Å/s and 3 Å/s respectively. Lift off process was performed with acetone followed by IPA cleaning.

### Electrical and Optical Measurement

All electrical measurement are carried out at room temperature under vacumm 10$^{-6}$ Torr without annealing samples inside the Janis cryogenic probe station.For photocurrent spectra measurument,FET devices were annealed for 2 h with 100 sccm Ar/H$_2$ (%15) flow at 200 ℃ to remove pmma residues and decrease contact resistance. Liqued Helium was connected to probe station by transfer line for low temperature PC spectra.

Acknowledgment. The authors would like to gratefully acknowledge financial support received from NSF through award: ECCS-1351424 (SK, FL), and partial support from the US Army grant, W911NF-10-2-0098, subaward 15-215456-03-00 (AV). This work was also partially supported (ADM, GG) by the LANL LDRD program (XW8V)). The work was conducted, in part, at the Center for Integrated Nanotechnologies (CINT), a U.S. Department of Energy, and Office of Basic Energy Sciences (OBES) user facility. ST acknowledges funding support provided by the U.S. Army Research Office through a MURI grant # W911NF-11-1-0362 and US

National Science Foundation (NSF) through grant # NSF-PIRE OISE-0968405. ST & KMD acknowledges funding support provided by Japan Society for the Promotion of Science (JSPS) through a fellowship (# L13521). MU would like to acknowledge partial support from National Science Foundation DMR CMMT Program (#1106214).

# Supporting Information-Chemical Vapor Deposition Synthesized Atomically-Thin Molybdenum Disulfide with Optoelectronic-Grade Crystalline Quality

Ismail Bilgin,[1,2] Fangze Liu,[1] Anthony Vargas,[1] Andrew Winchester,[3,4] Michael Ka Lun Man,[4] Moneesh Upmanyu,[5] Keshav Dani,[4] Gautam Gupta,[2] Saikat Talapatra,[3,4] Aditya D. Mohite*[2] and Swastik Kar*[1,6]

[1]Department of Physics, Northeastern University, Boston, Massachusetts 02115, United States
[2]Materials Synthesis and Integrated Devices, Los Alamos National Laboratory, Los Alamos, New Mexico 87545, United States
[3]Department of Physics, Southern Illinois University Carbondale, Carbondale, Illinois 62901, United States
[4]Femtosecond Spectroscopy Unit, Okinawa Institute of Science and Technology Graduate University, Onna, Okinawa 904-04905, Japan
[5]Department of Mechanical and Industrial Engineering, Northeastern University, Boston, Massachusetts 02115, United States
[6]George J. Kostas Research Institute for Homeland Security, Northeastern University, Burlington, Massachusetts 01803, United States

*Authors for Correspondence: amohite@lanl.gov and s.kar@neu.edu

## Sample Synthesis:

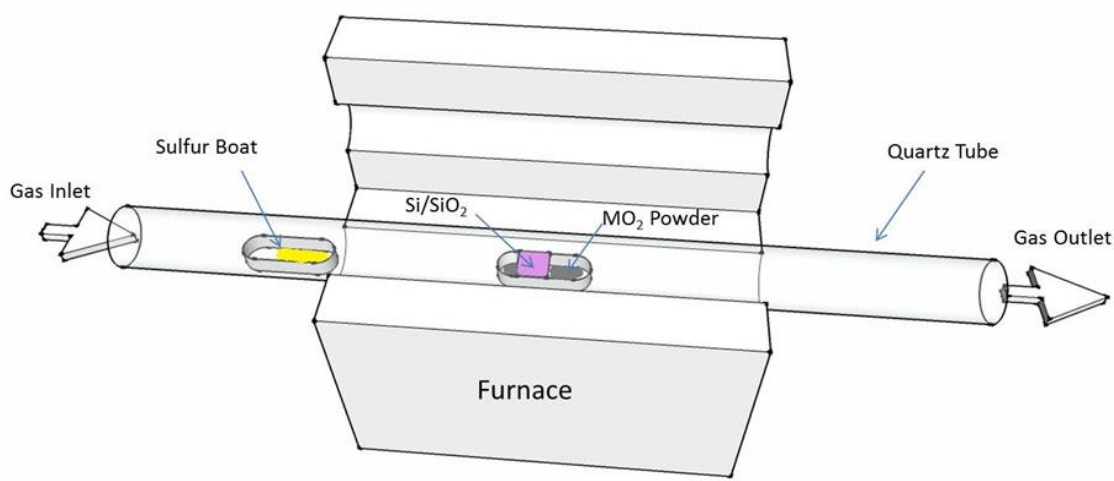

**Figure S1**: Schematic of the chemical vapor deposition (CVD) system for growing highly crystalline single layer MoS₂. Sulfur powder was placed upstream in the quartz tube in close proximity to the heating zone to allow the sulfur powder to vaporize (~110 °C) during the sulfurization of MoO₂ at 750 °C.

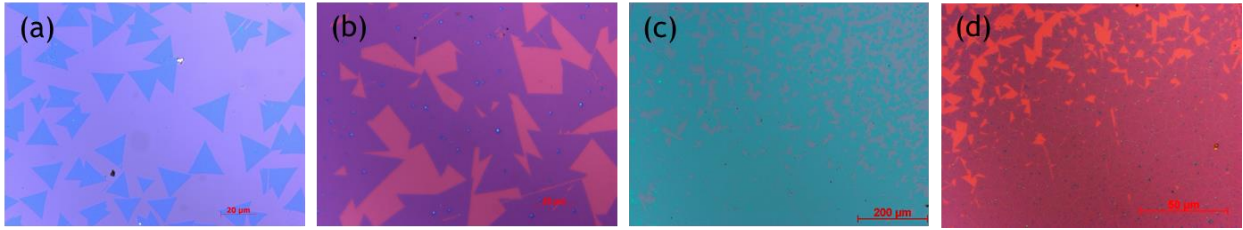

**Figure S2:** (a) Isolated single crystal monolayer samples. (b) With increasing growth duration, the crystals merge to form discontinuous polycrystalline samples. (c) and (d) Further increase of growth results in certain areas of the substrates being continuously covered with polycrystalline samples (bottom halves) while other parts (top halves) yet to complete the full-surface coverage.

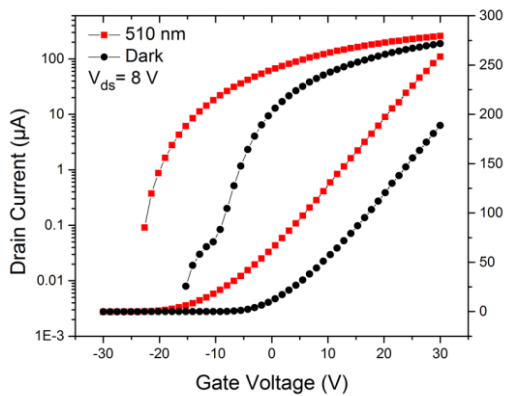

**Figure S3:** Transfer characteristic of the device in dark and under 510 nm light illumination at 8V bias voltage.

**Other Characterizations:**

**XPS:**

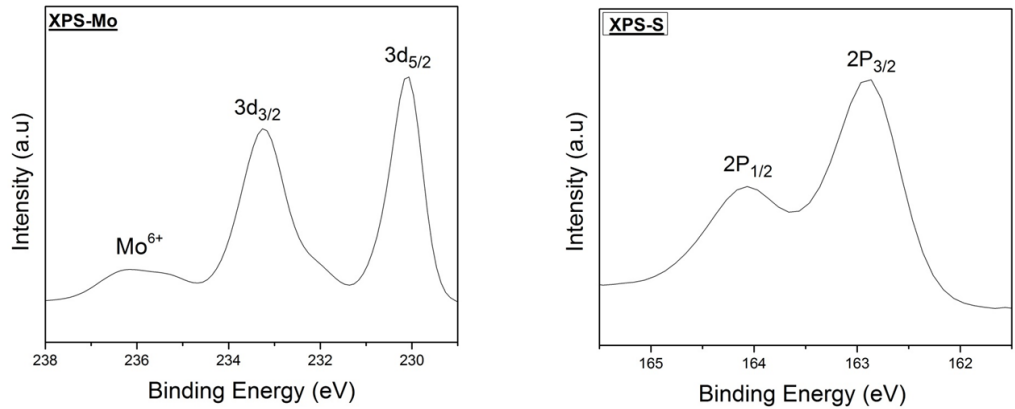

**Figure S4**: XPS spectra of monolayer $MoS_2$ for Mo 3d and S 2p. a) The Mo $3d_{3/2}$ and $3d_{5/2}$ peaks are at 233.26 and 230.06 respectively. $Mo^{6+}$ peak also observed at 236.16. b) S 2p binding energies are at 164.06 and 162.86. [1-3]

**EDX:**

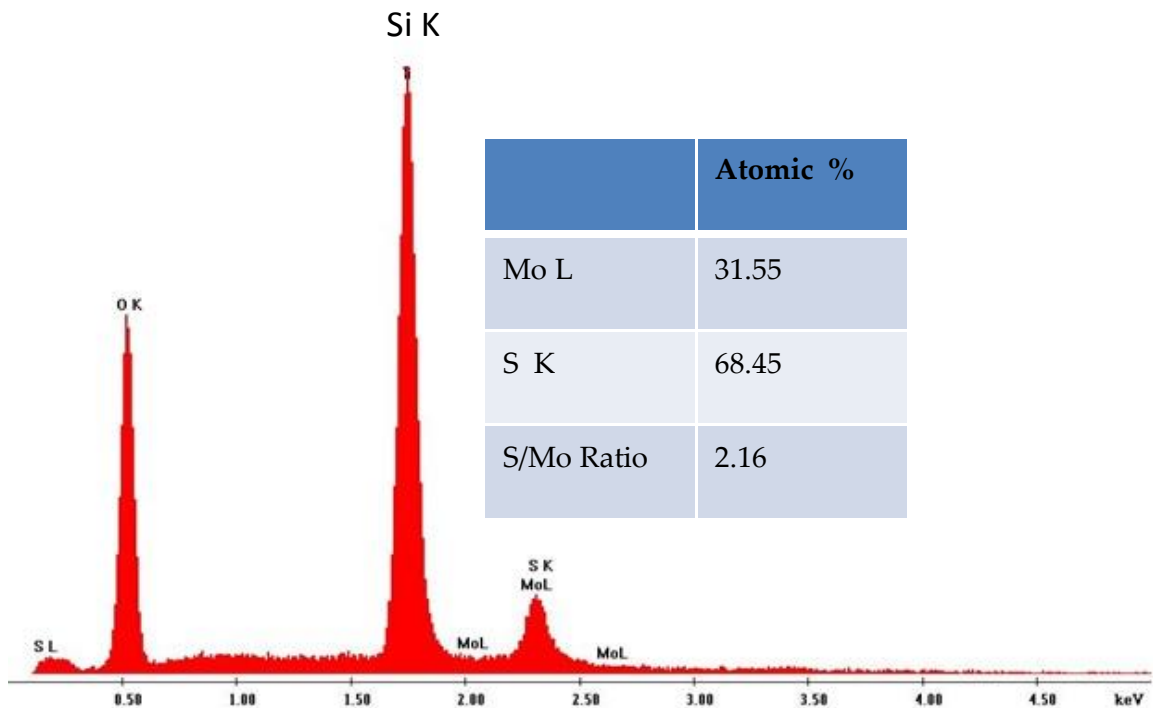

**Figure S5:** SEM-EDX spectrum of CVD grown single layer MoS2 on 300 nm Si/SiO₂. Inset shows atomic percentage of S and Mo atoms.

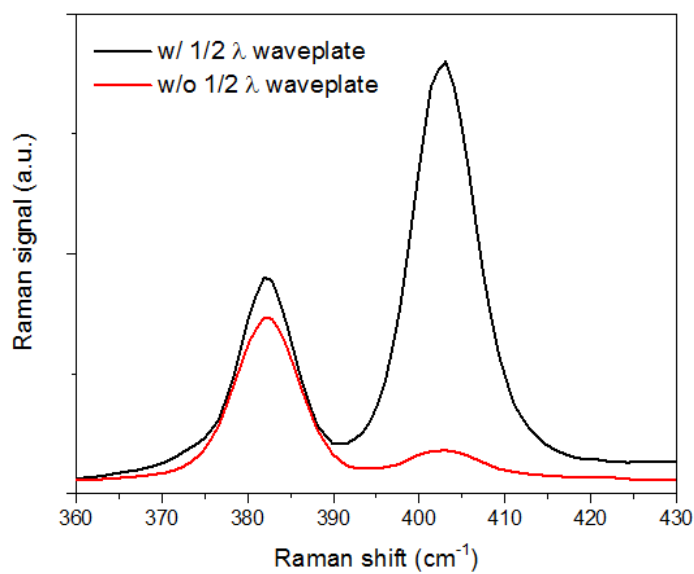

**Figure S6:** The $E^1_{2g}$ and $A_{1g}$ modes have both helicity[4] and linear polarization[5] dependence. In our Raman spectrometer, the incident laser is partially linear polarized. Therefore, the A_1g scattered light has the same polarization as the incident laser. The Raman has an optional ½ λ waveplate in front of the spectrometer. We tested the Raman spectra for both with and without the waveplate. The $A_{1g}$ peak shows strong polarization dependence compared to the $E^1_{2g}$ peak which shows weak dependence. This polarization dependence is not related to the sample quality.